\documentclass[]{article}

\usepackage{cite}
\usepackage{indentfirst}
\usepackage{newlfont}
\usepackage{amsthm}
\usepackage{amsmath}
\usepackage{latexsym}
\usepackage{amsmath, amssymb, amsfonts, mathrsfs, dsfont, mathtools, empheq}
\usepackage{amsthm}
\usepackage{epstopdf}
\usepackage{graphicx}

\usepackage{comment}
\usepackage{color}
\newcommand{\virg}[1]{``#1''}

\theoremstyle{remark}
\newtheorem{remark}{Remark}

\theoremstyle{plain}
\newtheorem{lemma}{Lemma}
\newtheorem{corollary}{Corollary}
\newtheorem{theorem}{Theorem}
\newtheorem{proposition}{Proposition}
\def\be{\begin{equation}}
\def\ee{\end{equation}}

\newcommand{\DI}{D}
\newcommand{\h}{h}
\newcommand{\J}{J}
\newcommand{\di}{d}
\newcommand{\R}{\mathbb R}
\newcommand{\E}{\mathbb E}
\newcommand{\1}{\mathds 1}
\newcommand{\N}{\mathbb N}
\newcommand{\DN}{\mathscr{D}_N}
\newcommand{\lga}{\log\gamma}

\newcommand{\eps}{\epsilon}

\title{Two populations mean-field monomer-dimer model}
\author{Diego Alberici, Emanuele Mingione}
\date{}

\begin{document}
\maketitle

\abstract
A two populations mean-field monomer-dimer model including both hard-core and attractive interactions between dimers is considered. The pressure density in the thermodynamic limit is proved to satisfy a three-dimensional variational principle. A detailed analysis is made in the limit in which one population is much smaller than the other and a ferromagnetic mean-field phase transition is found.

\section{Introduction}
Monomer-dimer models have been introduced in theoretical physics during the '70s to explain the absorption of diatomic molecules on a two-dimensional layer \cite{Rob}. Fundamental results were obtained by Heilmann and Lieb, who proved the absence of phase transitions \cite{HL} when only the hard-core interaction is taken into account, while the presence of an additional interaction coupling dimers can generate critical behaviours \cite{HLliquid}. Monomer-dimers  models have been source of a renewed interest in the last years in mathematical physics \cite{DG, Alb,GJL,AC}, condensed matter physics \cite{QDM} and in the applications to computer science \cite{ZM,KS} and social sciences \cite{BCSV,CVB}. The presence of an interaction beyond the hard-core one that couples different dimers is fundamental for the applications where phase transitions are observed \cite{BCSV,CVB}. Indeed in \cite{ACM,EPL, ACFM} the authors proved that a mean-field monomer-dimer model exhibits a ferromagnetic phase transition when a sufficiently strong interaction is introduced between pairs of dimers.

In this paper the investigation is extended to the case of a mean-field monomer-dimer model defined over two populations. This  multi-species framework  has been already  introduced  in the context of spin models  \cite{Mic1,BGG1,BCMT,Pa15} reveling interesting mathematical features. Multi-species monomer-dimer models are suitable to describe the experimental situation treated in \cite{BCSV,CVB}, where a mean-field type phase transition has been observed in the percentage of mixed marriages between native people and immigrants. The hard-core interaction between dimers naturally represents the monogamy constraint in marriages, while, as pointed out by the authors of \cite{BCSV}, an additional imitative interaction between individuals can be at the origin of the observed critical behaviour.

In this work we consider a mean-field model built on two populations $A$ and $B$ (e.g., the immigrants population and the local one) which takes into account both the imitative and the hard-core interactions. Dimers can be divided into three classes: type $A$ if they link two individuals in $A$,  type $B$ if they link two individuals in $B$ and type $AB$ if they link a mixed couple. The relative size of the two populations is fixed $N_A/N_B=\alpha/(1-\alpha)$. The energy contribution of  dimers is tuned  by a three dimensional vector $\h=(h_A,h_B, h_{AB})\in \R^3$ where
$h_{A}$ tunes the activity of a dimer of type $A$ and so on. Individuals have also  a certain propensity to imitate or counter-imitate the behaviour of the other individuals which is encoded in  an additional contribution to the energy tuned by a $3\times3$ real matrix $\J$. For example the entry $J_{AB}^{AB}$ couples dimers of type $AB$ with other dimers of the same type. The main result we obtain  is a representation of the pressure density in the thermodynamic limit in terms of a variational  problem in $\R^3$ for all the values of the  parameters $\h$ and $\J$ (see Theorem \ref{main} in section \ref{sec: model} for the precise statement). This result is applied in the  case where the only non-zero parameters contributing to the energy are $h_{AB}$ and $J_{AB}^{AB}$. As a  consequence the relevant degree of freedom of the model is the density of mixed dimers $d_{AB}$ and the above variational problem leads to a consistency equation of the type
\[ f_\alpha(d_{AB})\,=\,h_{AB}+J_{AB}^{AB}\,d_{AB}. \]
Its analytical properties are investigated in details for small $\alpha$: the mean-field critical exponent $1/2$ is rigorously found, consistently with the experimental situation described in \cite{BCSV,CVB}.

The paper is structured as follows.
In section \ref{sec: model} we introduce the  statistical mechanics  model with the basic definitions and we prove the main result: the thermodynamic limit of the pressure density  is expressed  as a three-dimensional variational problem, where the order parameters are the dimer densities $d_A,\,d_B$ internal to each population and the mixed dimer density $d_{AB}$.

In section \ref{sec: crit} we focus on three non-zero parameters, $\alpha,\,h_{AB},\,J_{AB}^{AB}$, and we study in detail the critical behaviour of the system when one population is much larger than the other ($\alpha\to0$), finding  a phase transition with standard mean-field exponents.

Finally in the Appendix we give an alternative proof for the existence of thermodynamic limit of the pressure density in the case $J=0$, $h_A + h_B \geq 2 h_{AB}$. This proof, which easily applies also to the standard single population case, uses a convexity inequality and is based on the Gaussian representation for the partition function \cite{ACMrand}.

\section{Model and main result} \label{sec: model}
Consider a system composed by $N$ sites divided into two populations of sizes $N_A$ and $N_B$ respectively, $N_A+N_B=N$.
We assume that the ratios $\alpha=N_A/N$ and $1-\alpha=N_B/N$ are fixed when the total size $N$ of the system varies.
A \textit{monomer-dimer configuration} can be identified with a set $\Delta$ of edges that satisfies a hard-core condition:
\be\label{def: md conf}
e=\{i,j\}\in\Delta \ ,\ e'=\{i',j'\}\in\Delta \quad\Rightarrow\quad e\cap e'=\emptyset
\ee

\begin{figure}[h]
\centering
\includegraphics[scale=0.5]{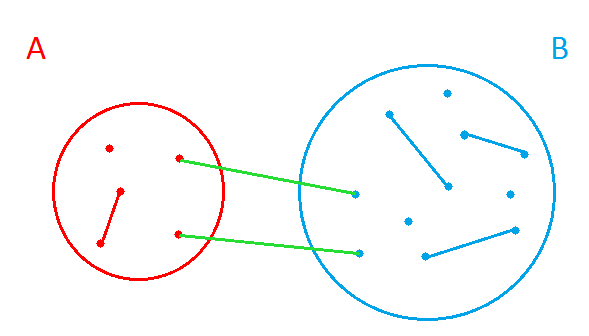}
\caption{A monomer-dimer configuration on two populations of sizes $N_A=5$, $N_B=11$. In this example there are $D_A=1$ dimers internal to population $A$, $D_B=3$ dimers internal to population $B$ and $D_{AB}=2$ mixed dimers.}
\label{fig: md}
\end{figure}

Given the configuration $\Delta$ (see Figure \ref{fig: md}), the edges in $\Delta$ are called dimers and they can be partitioned into three families: denote by $D_A$ the number of dimers having both endpoints in $A$, by $D_B$ the number of dimers having both endpoints in $B$ and by $D_{AB}$ the number of dimers having one endpoint in $A$ and the other one in $B$.
Monomers, namely sites free of dimers, can be partitioned into two families: denote by $M_A,\,M_B$ the number of monomers in $A,\,B$ respectively.
Observe that
\be\label{eq: hardcore}
 2\,D_A+D_{AB}+M_A = N_A \quad,\quad 2\,D_B+D_{AB}+M_B = N_B
\ee

We denote by $\DN$ the set of all possible monomer-dimer configurations on $N$ sites. 
For a given configuration $\Delta\,\in\, \DN$,  $\DI$ denotes the vector of the cardinalities of the three families of dimers
\be
\DI \,:=\, \left(\begin{array}{l} D_A \\[2pt] D_B \\[2pt] D_{AB} \end{array}\right) \;.
\ee
while
\be
|D| \,:=\, D_A+D_B+D_{AB}
\ee
represents the total number of dimers.
The  Hamiltonian function is defined as
\be \label{eq: hamiltonian}
H_N(D) \,=\, - \h\,\cdot\,\DI - \frac{1}{2 N} \,\J \DI\,\cdot\, \DI
\ee
where $\cdot$ denotes the standard scalar product in $\R^3$, the dimer vector field $h$ tunes the activity of dimers while the coupling matrix $\J$ tunes the interaction between sites according to the types of dimers they host: 
\be
\h \,=\, \left(\begin{array}{l} h_A \\[2pt] h_B \\[2pt] h_{AB} \end{array}\right)
\quad
\J \,=\, \left(\begin{array}{lll} J_A^A & J_A^B & J_A^{AB} \\[2pt] J_B^A & J_B^B & J_B^{AB} \\[2pt] J_{AB}^A & J_{AB}^B & J_{AB}^{AB} \end{array}\right) \;.
\ee
The  partition function of the model is
\be\label{partitionf}
Z_N \,\equiv\,Z_N(\h,\J,\alpha)\,=\, \sum_{\Delta\in\DN} \, N^{-|D|}\, e^{-H_N(D)}
\ee
where the term $N^{-|D|}$ is necessary to ensure a well defined thermodynamic limit of the model.
Given $f\!:\DN \to \R $ we call expected value of $f$ with respect to the Gibbs measure the quantity
\be
\label{Gibbs measure}
\langle\,f\,\rangle_N\,:=\, \dfrac{1}{Z_N} \sum_{\Delta\in\DN} \, N^{-|D|} e^{-H_N(D)}\,f(\Delta)
\ee
where $H_N$ is the Hamiltonian function \eqref{eq: hamiltonian}.\\

Let us introduce the definitions needed to state our main result.
Denote by $\Omega_\alpha$ the set of $\di=(d_A,d_B,d_{AB})^T\in(\R_+)^3$ such that
\be
2d_A+d_{AB}\leq\alpha\,,\ 2 d_B+d_{AB}\leq1-\alpha \;.
\ee
The above constraints on the vector $d$ reflect the hard-core relations \eqref{eq: hardcore}. Set
\be \label{eq: gamma}
\gamma(x) :=\, \exp(x\log x-x)\ , \quad x\geq0
\ee
and define the following functions
\be \label{eq: s} \begin{split}
s(\di;\alpha) :=\; & \lga(\alpha) + \lga(1-\alpha) -\lga(\alpha-2 d_A-d_{AB}) \;+\\[2pt]
& -\lga(1-\alpha-2 d_B-d_{AB}) -\lga(d_A) -\lga(d_B) \;+\\[2pt]
& -\lga(d_{AB}) - d_A\log2 - d_B\log2
\end{split} \ee
\be \label{eq: eps}
\eps(\di;\h,\J) :=\, - h\,\cdot\, d - \frac{1}{2}\,Jd\,\cdot\,d  \\[2pt]
\ee
\be  \label{eq: psi}
\psi(\di;\h,\J,\alpha) :=\, s(d;\alpha) - \eps(d;h,J) \;. \\[4pt]
\ee
The functions $\psi,s,\eps$ represent respectively the variational pressure, entropy and energy densities.

\begin{theorem}\label{main}
For all $\alpha\in(0,1)$, $h\in\R^3$ and $J\in\R^{3\times3}$, there exists
\be \label{eq: p}
\lim_{N\to\infty}\frac{1}{N}\log Z_N(\h,\J,\alpha) \,=\, \max_{\di\in\Omega_\alpha} \psi(\di;\h,\J,\alpha) \,=: p(\h,\J,\alpha)
\ee
The function $\psi(\di;\h,\J,\alpha)$ attains its maximum in at least one point $\di^*=\di^*(\h,\J,\alpha)\in\Omega_\alpha$ which solves the following fixed point system:
\be\label{eq: system}
\begin{cases}\, d_A=\frac{w_A}{2}\,m_A^2 \\ d_B=\frac{w_B}{2}\,m_B^2 \\ d_{AB}=w_{AB}\,m_A\,m_B \end{cases}
\ee
where we denote
\be
m_A= \alpha-2d_A-d_{AB}\ ,\quad m_B= 1-\alpha-2d_B-d_{AB}\ ,
\ee
\be
w_A= e^{h_A+J_Ad}\ ,\quad w_B= e^{h_B+J_Bd}\ ,\quad w_{AB}= e^{h_{AB}+J_{AB}d}\ .
\ee
At $\J=0$ the system \eqref{eq: system} has a unique solution $\di^*=g(h,\alpha)\in\Omega_\alpha$ which is an analytic function of the parameters $\h,\alpha$. Clearly at any $J$ the system \eqref{eq: system} rewrites as
\be\label{eq: system1}
\di \,=\, g(\h+\J\di\,,\,\alpha) \ .
\ee
Provided that $\di^*$ is differentiable, $\nabla_{\!\h\,} p = \di^*$ hence there exists
\be
\lim_{N\to\infty}\frac{1}{N}\left\langle \,\DI\, \right\rangle_N \,=\, \di^* \;.
\ee
\end{theorem}


\proof
The number of configurations $\Delta\in\DN$ with given cardinalities $D_A$, $D_B$, $D_{AB}$ can be computed by a standard combinatorial argument. Therefore the partition function rewrites as
\be
Z_N \,=\, \sum_{D_A=0}^{N_A/2}\, \sum_{D_B=0}^{N_B/2} \sum_{D_{AB}=0}^{\,(N_A-2D_A)\land(N_B-2D_B)} \phi_N(\DI)\,e^{-H_N(\DI)}
\ee
with
\be
\phi_N(\DI) := \frac{N_A!\,N_B!}{(N_A-2D_A-D_{AB})!\,(N_B-2D_B-D_{AB})!\,D_A!\,D_B!\,D_{AB}!\,2^{D_A}\,2^{D_B}}
\ee
In order to simplify the computations, we approximate the factorial by the continuous function $\gamma$ defined in \eqref{eq: gamma}.
We denote by $\tilde\phi_N$ the function obtained from $\phi_N$ by substituting any factorial $n!$ with $\gamma(n)$, then we denote by $\tilde Z_N$ the partition function obtained from $Z_N$ by substituting $\phi_N$ with $\tilde\phi_N$.
The Stirling approximation and elementary computations give the following properties of $\gamma$:
\begin{itemize}
\item[i.] $1\lor\sqrt{2\pi n} \,\leq\, n!/\gamma(n) \,\leq\, 1\lor e^{\frac{1}{12}}\sqrt{2\pi n} \quad\forall n\in\N$
\item[ii.] $\frac{d}{dx}\log\gamma(x) = \log x\,,\quad \log\gamma(x)$ is convex
\item[iii.] $\frac{1}{N}\log\gamma(Nx)=\log\gamma(x)+x\,\log N$
\end{itemize}
By i. it follows that
\be\label{eq: tilde}
\frac{1}{N}\log Z_N \,=\, \frac{1}{N}\log\tilde Z_N \,+\, \cal O\left(\frac{\log N}{N}\right) \;,
\ee
by a standard argument
\be\label{eq: max}
\frac{1}{N}\log\tilde Z_N \,=\, \max_{\DI\in N\Omega_\alpha} \frac{1}{N}\left(\log\tilde\phi_N(\DI)-H_N(\DI)\right) \,+\, \cal O\left(\frac{\log N}{N}\right)
\ee
and using iii. a direct computation shows that for every $N\in\N$
\be
\frac{1}{N}\left(\log\tilde\phi_N(N\di)-H_N(N\di)\right) \,=\, \psi(\di;\h,\J,\alpha) \,,\quad \di\in\Omega_\alpha \;.
\ee
Therefore there exists
\[ \lim_{N\to\infty} \frac{1}{N}\log Z_N = \max_{\di\in\Omega_\alpha}\psi(\di;\h,\J,\alpha) \;.\]
Using ii. one can easily compute
\be
\nabla_{\!\di\,}s \,=\, \left(\, \log\frac{m_A^2}{2d_A} \ ,\ \log\frac{m_B^2}{2d_B}  \ ,\ \log\frac{m_Am_B}{d_{AB}} \,\right)
\ee
\be
-\nabla_{\!\di\,}\eps \,=\, \left(\,  h_A+\J_A\cdot\di \ ,\ h_B+\J_B\cdot\di \ ,\ h_{AB}+\J_{AB}\cdot\di \,\right)
\ee
therefore
\[ \nabla_{\!\di\,}\psi(\di;\h,\J,\alpha) \,=\, 0 \ \Leftrightarrow\ \di \text{ is a solution of \eqref{eq: system}} \;.\]
The first derivatives of $p(\h,\J,\alpha)=\psi(\di^*(\h,\J,\alpha);\h,\J,\alpha)$ can be easily computed since $\nabla_{\!\di\,}\psi(\di^*;\h,\J,\alpha)=0$.



\section{The limit $\alpha\,\to\, 0$} \label{sec: crit}
In this section we choose a particular framework that simplifies the mathematical treatment of the problem and allows a detailed analysis of the thermodynamic properties of the system.
The most peculiar parameters of the model are $h_{AB}$ and $J_{AB}^{AB}$, describing respectively the $AB$-dimer field and the interaction between couples of $AB$-dimers, indeed they have no correspondence in a bipopulated Ising model \cite{Mic1}.
Moreover we focus on the case where one population is much smaller than the other ($\alpha\to0$), since it is interesting for the social applications \cite{BCSV}.
Thus in this section we set $h_A=h_B=0$, $J_A^A=J_A^B=J_B^A=J_{A}^{AB}=J_{AB}^A=J_{B}^{AB}=J_{AB}^B=0$ and we consider only the remaining coefficients $h_{AB}$ and $J_{AB}^{AB}$. From now on, with a slight abuse of notation, we will denote
\[ h := h_{AB}\ ,\quad J := J_{AB}^{AB} >0 \]
and the mixed dimer density
\[ d := d_{AB} = \frac{D_{AB}}{N} \in [0,\alpha] \]



\begin{figure}[h]
\centering
\includegraphics[scale=0.35]{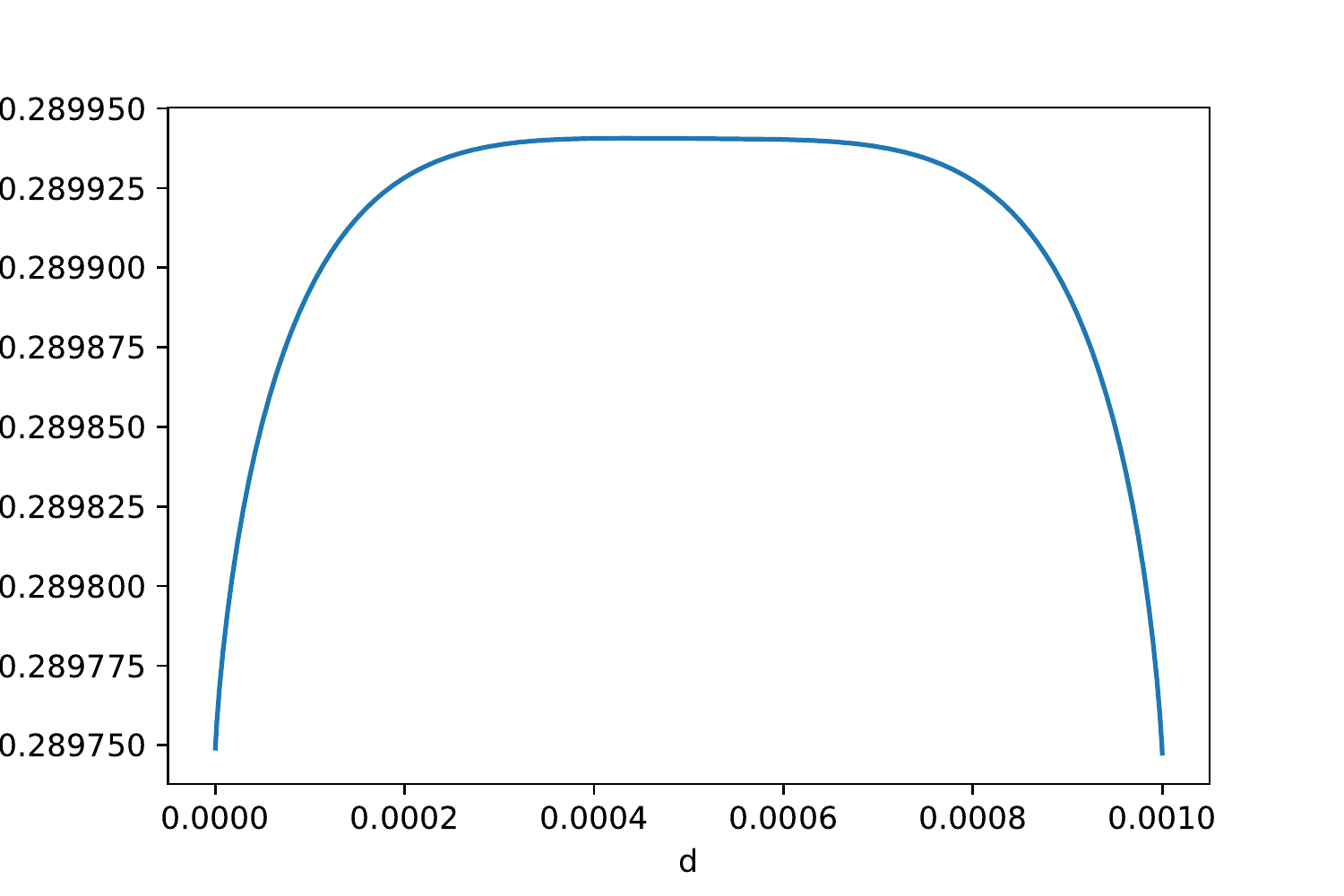}
\includegraphics[scale=0.35]{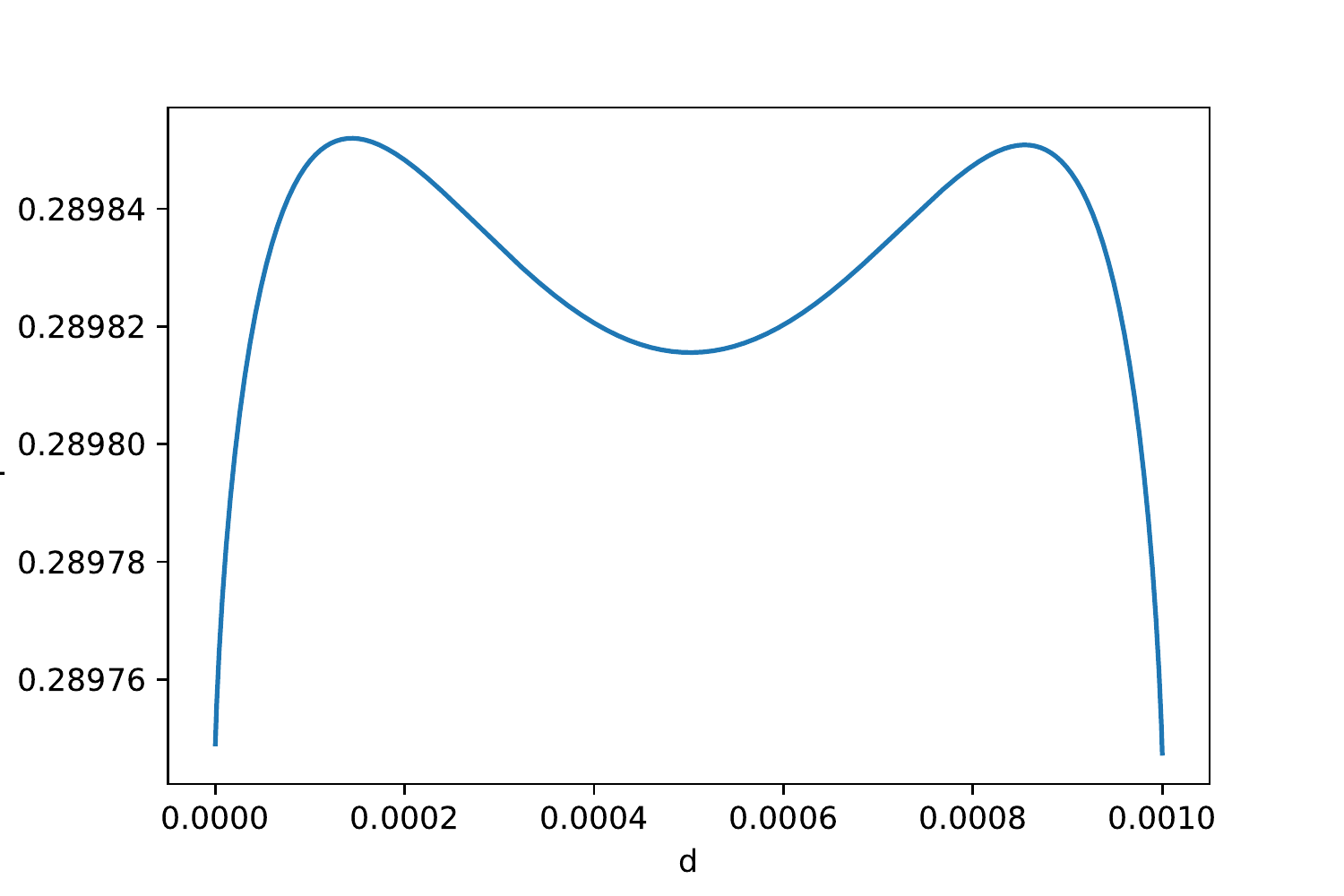}
\caption{Plots of the variational pressure $\psi_1$ versus $d$, for $\alpha=10^{-3}$ and different values of the parameters: at the critical point $J=J_c$, $h=h_c$ on the left-hand side, at the point $J=J_c+10^3$, $h=h_c-d_c\,(J-J_c)$ on the right-hand side. Moving from the critical point along a suitable curve, the global maximum points of $\psi_1$, that by \eqref{eq: p2} identify the phases of the system, pass from one to two.}
\label{fig: p}
\end{figure}

In this framework the degrees of freedom of the variational problem \eqref{eq: p} reduces from three to one, since $d_A,d_B$ are explicit functions of $d_{AB}\equiv d$ as can be easily observed by looking to the consistency equation \eqref{eq: system}.
Precisely, by setting $x_\alpha(d):=m_A=\sqrt{2d_A}\,$, $y_\alpha(d):=m_B=\sqrt{2d_B}$
one can easily see that $x_\alpha(d),\,y_\alpha(d)$ are the positive solutions of the following quadratic equations respectively
\be
x^2 + x -(\alpha-d) =0 \quad,\quad y^2+y-(1-\alpha-d) =0
\ee
namely
\be
x_\alpha(d) = \frac{-1+\sqrt{1+4(\alpha-d)}}{2} \quad,\quad y_\alpha(d) = \frac{-1+\sqrt{1+4(1-\alpha-d)}}{2} \;.
\ee
Then one can easily prove from Theorem \ref{main} that
\be \label{eq: p2}
p(h,J,\alpha) \,=\, \max_{d\,\in\,(0,\alpha)} \psi_1(d;h,J,\alpha)
\ee
where $\psi_1$ coincides with the function $\psi$ defined by equation \eqref{eq: psi} evaluated at
\be
\left(\begin{array}{l} d_A \\[2pt] d_B \\[2pt] d_{AB} \end{array}\right) = \left(\begin{array}{l} x_{\alpha}(d)^2/2 \\[2pt] y_{\alpha}(d)^2/2 \\[2pt] d \end{array}\right) \;.
\ee
Any solution ${d}^*={d}^*(h,J,\alpha)$ of the one-dimensional variational problem \eqref{eq: p2} satisfies the fixed point equation
\be\label{eq: d cons}
d\,=\, \exp(h+Jd) \,x_\alpha(d)\, y_\alpha(d)
\ee
It is convenient to set $f_\alpha(d):=\log d-\log x_\alpha(d)-\log y_\alpha(d)$ and rewrite equation \eqref{eq: d cons} as $f_\alpha(d)=h+Jd\,$. Fix $\alpha\in(0,1)$. $f_\alpha$ is the inverse function of a sigmoid function\footnote{It is easy to check that $f_\alpha(d)\to-\infty$ as $d\searrow0$, $f_\alpha(d)\to\infty$ as $d\nearrow\alpha$, $f_\alpha'>0$, $f_\alpha''$ vanishes exactly once.}. Therefore the point $(d_c,h_c,J_c)$ such that $f''_\alpha(d_c)=0$, $f'_\alpha(d_c) = J_c$, $f_\alpha(d_c)=h_c+J_c\,d_c$ is the critical point of the system, where the density ${d}^*$ branches from one to two values (see Figure \ref{fig: p}).

For small values of $\alpha$, the following estimates for the critical point can be obtained by expanding $f_\alpha(d)$ as $\alpha\to0$:
\be
 d_c(\alpha) = \frac{\alpha}{2} + \mathcal{O}(\alpha^3)
\ee
\be
J_c(\alpha) = \frac{4}{\alpha} + \mathcal{O}(\alpha)
\ee
\be
h_c(\alpha) = -2-\log\frac{\sqrt5-1}{2} + \mathcal{O}(\alpha)
\ee

Fixing $\alpha$ close to zero and moving the parameters $(h,J)$ towards their critical values, along the half line $h-h_c(\alpha)=-d_c(\alpha)\,\big(J-J_c(\alpha)\big)$, $J\geq J_c$, the mixed dimer density $d^*(h,J,\alpha)$ exhibits the following critical behaviour:
\be\label{radice}
{d}^*(h,J,\alpha)-d_c(\alpha) \,=\, C(\alpha)\,\sqrt{J-J_c(\alpha)} \,+\, \mathcal{O}\Big((J-J_c(\alpha))^{3/2}\Big)
\ee
with $C(\alpha)=\sqrt{\frac{3}{16}\alpha^{3}+\mathcal{O}(\alpha^{6})}$. This fact can be proven using the Taylor expansion of $f_\alpha(d)$ around $d=d_c(\alpha)$ up to the third order.


\begin{remark}
It is remarkable  that our model is  in good agreement with the experimental results in \cite{BCSV} where the authors find that the fraction of mixed marriage over total number of marriages
\be\label{mix}
d_{mix}=\lim_{N\to\infty}\left\langle\frac{D_{AB}}{|D|}\right\rangle
\ee
undergoes a mean-field like phase transition for small values of $\alpha$. More precisely they obtain that a function of the type
\be\label{experim}
d_{mix}(\alpha)\,=\,C \, \sqrt{\alpha-\alpha_c}\;,\ \alpha >\alpha_c\approx 0.005\,,
\ee
is a very good fit for the experimental values of $d_{mix}$ versus $\alpha$.

The critical behaviour \eqref{experim} can be predicted by the model presented in this section, with coupling $J = \alpha\,(1-\alpha)\,J'$, $J'\gg1$. Indeed, for fixed $J'\gg1$, the critical point of the system is given by $(d_c,h_c,\alpha_c)$, where

\be\alpha_c=\frac{2}{\sqrt{J'}}\,+\,\mathcal{O}(\frac{1}{J'})
\ee
\be h_c=-2-\log\frac{\sqrt5-1}{2}\,+\,\mathcal{O}(\frac{1}{\sqrt{J'}})
\ee
\be d_c=\frac{1}{\sqrt{J'}}\,+\,\mathcal{O}(\frac{1}{J'^{\,3/2}})
\ee
and the critical behaviour of $d_{mix}$ as $\alpha\to\alpha_c$, $h=h_c-d_c\,(\alpha-\alpha_c)$, is the following:
\be\label{crti}
d_{mix}-(d_{mix})_c \,=\, C(J')\,\sqrt{\alpha-\alpha_c} \,+\, \mathcal{O}\big((\alpha-\alpha_c)^{3/2}\big)
\ee
where
\[ (d_{mix})_c \,=\, \frac{d_c}{\frac{1}{2}x(d_c)^2+\frac{1}{2}y(d_c)^2+d_c} =
\frac{2}{3-\sqrt5}\,\alpha_c + \mathcal{O}(\frac{1}{J'}) \;.\]
\end{remark}

\begin{remark}
Equation \eqref{crti} is a consequence of the fact that at the critical point the lowest order non vanishing derivative of the variational pressure $\psi_1$ in \eqref{eq: p2} is the fourth one. This fact suggests that the fluctuations of the order parameter at the critical point follows the standard mean field theory \cite{ellis1978limit, ACFM}. From the above considerations we expect the fluctuations scale as $N^{3/4}$ and converge to a quartic exponential distribution agreement with the experimental results in \cite{CVB}.
\end{remark}

\textbf{Acknowledgment:} We thank Pierluigi Contucci for bringing the problem to our attention  and
we acknowledge  financial support by GNFM-INdAM Progetto Giovani 2017.

\section*{Appendix}
Here we give a directed proof of the existence of the thermodynamic limit for the pressure density in the particular case
\be \label{eq: W>0}
J=0 \ ,\quad W=\begin{pmatrix}w_A & w_{AB}\\ w_{AB} & w_B\end{pmatrix} = \begin{pmatrix}e^{h_A} & e^{h_{AB}}\\ e^{h_{AB}} & e^{h_B}\end{pmatrix} >0 \ .
\ee
where $W>0$ means that the matrix $W$ is positive definite.
This proof is independent from Theorem \ref{main} and the strategy follows a basic idea introduced in \cite{GT} in the context of Spin Glass Theory. In this case the partition function \eqref{partitionf} admits a representation in terms of Gaussian moments:
\be \label{eq: Gauss}
Z_N \,=\, \sum_{\Delta\in\DN} \left(\frac{w_A}{N}\right)^{D_A}\left(\frac{w_B}{N}\right)^{D_B}\left(\frac{w_{AB}}{N}\right)^{D_{AB}} \,=\, \E\left[(1+\xi_A)^{N_A}(1+\xi_B)^{N_B}\right]
\ee
where $\xi =(\xi_A,\xi_B)$ is a centred Gaussian vector of covariance matrix $\frac{1}{N}W$ (the hypothesis of positive definiteness is crucial).
The representation \eqref{eq: Gauss} is based on the Isserlis-Wick formula, see \cite{ACMrand} (Proposition 2.2) for the proof.

Now consider the set $Q=\{\xi\in\R^2 \,:\, 1+\xi_A>0,\,1+\xi_B>0\}\,$ and define a modified partition function
\be \label{eq: Gauss1}
Z_N^* =\, \E\left[(1+\xi_A)^{N_A}(1+\xi_B)^{N_B}\,\1_Q(\xi) \right]
\ee
$Z_N^*$ rewrites as an integral over $\xi\in Q$ with integrand function proportional to  $\exp(N\,f(\xi) )$ where
\[ f(\xi)= -\frac{1}{2}\langle W^{-1}\xi,\xi\rangle+\alpha\log|1+\xi_A|+(1-\alpha)\log|1+\xi_B| \]
Since $f$ approaches its global maximum on $\R^2$ only for $\xi_A\geq0,\,\xi_B\geq0$, standard Laplace type estimates implies that
\be  \label{eq: ZZ*}
\frac{Z_N}{Z_N^*} \rightarrow 1 \quad\text{as }N\to\infty\ .
\ee
Hence we can restrict our attention to the sequence $\log Z_N^*$, $N\in\N$.
We claim that
\begin{proposition} \label{prop: superadd}
For every $N_1,N_2,N\in\N$ such that $N=N_1+N_2$, it holds
\be
Z_{N_1}^*\,Z_{N_2}^* \,\leq\, Z_N^* \;.
\ee
\end{proposition}
Then the sequence $\log Z_N^*$ is super-additive and the \virg{monotonic} convergence of the pressure density will follow immediately by Fekete's lemma and equation \eqref{eq: ZZ*}:
\begin{corollary}
Under the hypothesis \eqref{eq: W>0}, there exists
\be
\lim_{N\to\infty} \frac{1}{N}\log Z_N \,=\, \sup_{N} \frac{1}{N}\log Z_N^*
\ee
\end{corollary}
Only the proposition \ref{prop: superadd} remains to be proven.

\proof[Proof of the proposition \ref{prop: superadd}]
The strategy for the proof follows the basic ideas introduced in \cite{GT}
for mean field spin models. For a fixed $N$ consider two integers  $N_1,N_2$, such that $N=N_1+N_2$ and set
\[ \gamma=N_1/N\,,\ 1-\gamma=N_2/N \;,\]
We decompose each of the two parts of the system $N_1,N_2$  in two populations $A,B$ according to the fixed ratio $\alpha$, namely according to the relation
\[ N_i = \alpha N_i + (1-\alpha) N_i =: N_{iA}+N_{iB} \,,\quad i=1,2 \]
Now we introduce two \textit{independent} centred Gaussian vectors:
\[ \xi_i = (\xi_{iA}\,,\,\xi_{iB})\,\ \text{with covariance matrix }\frac{1}{N_i}\,W \,,\quad i=1,2 \]
and we prove the following lemmas.

\begin{lemma} \label{lem: xi'}
\[ \gamma\,\xi_1+(1-\gamma)\,\xi_2 \,\overset{d}{=}\, \xi \]
\end{lemma}
\proof
Since $\xi_1,\xi_2$ are independent centred Gaussian vectors, $\xi':=\gamma\,\xi_1+(1-\gamma)\,\xi_2$ is a centred Gaussian vector. Its covariance matrix is:
\[ \gamma^2\,\frac{W}{N_1}+(1-\gamma)^2\,\frac{W}{N_2} \,=\, \gamma\,\frac{W}{N}+(1-\gamma)\,\frac{W}{N} \,=\, \frac{W}{N} \ ,\]
the same of $\xi$.
\endproof

\begin{lemma} \label{lem: ineq}
\[ (1+x)^\gamma\,(1+y)^{1-\gamma} \leq 1+\gamma x+(1-\gamma)y \quad\forall\,x>-1,\,y>-1,\gamma\in(0,1)  \]
\end{lemma}
\proof
Consider the function $f(x,y)=(1+x)^\gamma\,(1+y)^{1-\gamma}$ and its Taylor polynomial of first order at $(0,0)$, $P(x,y)=1+\gamma x+(1-\gamma)y\,$. The Hessian matrix of $f$ is negative defined for $x>-1,\,y>-1$ (it has zero determinant and negative trace), hence $f(x,y)\leq P(x,y)\,$.
\endproof

Finally the proof of proposition \ref{prop: superadd} follows easily using  the independence of $\xi_1,\,\xi_2$, lemma \ref{lem: ineq} and lemma \ref{lem: xi'} .
\endproof

\endproof

\end{document}